\documentclass{article}

\usepackage{arxiv}

\usepackage[utf8]{inputenc} 
\usepackage[T1]{fontenc}    
\usepackage{hyperref}       
\usepackage{url}            
\usepackage{booktabs}       
\usepackage{amsfonts}       
\usepackage{nicefrac}       
\usepackage{microtype}      
\usepackage{lipsum}
\usepackage{graphicx}
\graphicspath{ {./images/} }
\usepackage{natbib}
\usepackage{caption}

\title{Error in ERA5 2m Temperature identified using GraphCast}

\author{
 Hannah M. Christensen, Jack Barker, and Bobby Antonio \\
  Department of Physics, University of Oxford,\\
  Oxford, U.K.\\
  \texttt{hannah.christensen@physics.ox.ac.uk} \\
   \And
 Massimo Bonavita, Mohamed Dahoui, and Patricia de Rosnay \\
  European Centre for Medium-Range Weather Forecasts, \\
  Reading, U. K. \\
}

\begin{document}
\maketitle
\begin{abstract}
Reanalyses such as ERA5 have long been foundational for weather and climate science. They have also found a new use case, as training and verification data for machine-learnt weather prediction (MLWP) models. Here we compare short-lead time (6h) forecasts from the MLWP model GraphCast against ERA5. In doing so, we identify a recurrent, spatially coherent error in 2m Temperature centred on the Ethiopian Highlands, that occurs predominantly at 0600 UTC. We show that these error events are not an error in the forecast from GraphCast, but are in fact an error in ERA5, and are also present in the ECMWF operational analysis. They arise from the 2D optimal interpolation procedure, when surface reports are assimilated that are temporally displaced compared to the background forecast. This produces spuriously warm analysis increments over Ethiopia on approximately 7\% of dates at 0600 UTC across the reanalysis record. The spread from the ensemble of data assimilation partially flags these cases but is underdispersive. We assess the impact on GraphCast, which was trained on ERA5. While GraphCast can largely ignore these unphysical error events, a small systematic degradation in forecast skill over the region is observed. We discuss implications for using reanalysis as truth in machine learning training and verification, and recommend simple changes to reduce such artefacts in future analyses. 
\end{abstract}


\section{Introduction}\label{intro}

The ERA5 reanalysis, produced by the European Centre for Medium-Range Weather Forecasts (ECMWF), is a globally consistent, high-resolution reconstruction of past atmospheric, land, and ocean surface conditions from 1950 to present \citep{hersbach_era5_2020}. It is one of the most widely used climate datasets: the ERA5 reference paper \citep{hersbach_era5_2020} has received around 4000 citations a year since publication. ERA5 is used for a diverse range of applications: it is used to assess the severity of climate change by revealing long-term shifts in temperature, precipitation, and extreme events \citep{ipcc2021ar6}; it advances physical climate science through analysis of large-scale circulation patterns, variability, and teleconnections \citep[e.g.][]{oreilly2024stormtrack,kang2024qbo}; it is used to provide initial conditions for model simulations \citep{roberts2025highresmip2} and further serves as a benchmark, allowing comparison of model predictions against an observationally constrained dataset \citep[e.g.][]{strommen2025gulfstream}. 

Recent years have seen a new use case for ERA5 data: its use as training data for Machine-Learnt Weather Prediction (MLWP) models \citep{rasp2020weatherbench,rasp2024weatherbench2}.  These machine-learnt emulators are now positioned as powerful substitutes for physics-based numerical weather prediction (NWP) models \citep{Bi2023,lam_learning_2023}. MLWP models can produce forecasts up to 1,000 times faster compared to NWP \citep{Pathak2022}, and they also show improved forecast skill compared to their dynamical counterparts \citep{lam_learning_2023, Kochkov2024, Price2024, Alet2025, YuvalPrecipNeuralGCM2025}. Arguably, it is the availability of ERA5 --- including its spatio-temporal coverage, comprehensive set of variables, and open source licence --- that has made this Machine Learning revolution for weather prediction possible \citep{rasp2024weatherbench2}.

While initial evaluation of MLWP models focused on their skill compared to conventional numerical weather prediction (NWP) models, subsequent studies have considered their physical behaviour. This includes whether they emulate well-known physical relationships such as geostrophic balance~\citep{bonavita_limitations_2024} and the Matsuno–Gill response to tropical heating perturbations~\citep{hakim_dynamical_2024}. Other studies have considered Kelvin and Rossby waves~\citep{jalan_intraseasonal_2025}, teleconnections~\citep{diao_assessing_2025}, extreme events~\citep{bano-medina_are_2025, olivetti_data-driven_2024, meng_deep_2025}, and error growth~\citep{selz_can_2023}. 

One approach to understand the behaviour of a model better is to study the errors it makes. For example, the Transpose-AMIP experiment \citep{williams2013} used a reanalysis product to initialise climate models. Comparing their forecast errors over one to two day lead times highlighted fast growing model errors that could be attributed to the parametrised physics. Similarly, \citet{rodwell2007} used the error in short-term forecasts from within the data assimilation cycle to rule out certain parameter perturbations as unphysical.

The original goal of our study was to analyse errors made by a MLWP model, GraphCast \citep{lam_learning_2023}, over very short lead times to learn about the physical consistency of the model. However, in studying the characteristics of fast errors made with GraphCast, we identified a curious error pattern over the Ethiopian Highlands: on 7\% of days, GraphCast forecasts validating at 06:00UTC have surface temperature fields that are 5K colder than ERA5.  We find that this error is not due to GraphCast, but is rather a feature of the verification dataset, ERA5. It arises because a cluster of surface observations in this region are provided at non-synoptic times. This leads to a large systematic offset between the observations and background forecast. The observations involved are usually rejected by the quality control threshold; when they are included, they lead to a jump in the analysis of the observed magnitude. The presence of this error in ERA5 provides a novel test for the physical behaviour of GraphCast and other MLWP models trained using ERA5: can they learn to ignore this clearly unphysical feature in the training data?

This paper is structured as follows. In Section~\ref{sec:methods} we present the ERA5 dataset and the GraphCast model and describe the error database we constructed for our analysis. In Section~\ref{sec:error} we present the error and its characteristics. In Section~\ref{sec:cause} we discuss the cause of the error in ERA5. In Section~\ref{sec:implications} we discuss the implications for GraphCast forecasts, since this model (like many MLWP models) is trained using ERA5 data. Finally in Section~\ref{sec:concs} we reflect on the implications of the paper, and present some conclusions.

\section{Methods}\label{sec:methods}

GraphCast is a graph neural network-based model developed by Google DeepMind for medium-range weather forecasting~\cite{lam_learning_2023}. It evolves the 3D global state of atmospheric temperature, winds, humidity and geopotential on 37 pressure levels, as well as the state of five surface fields. It can produce a 10-day forecast in under a minute on a single Google Tensor Processing Unit (TPU). Predictions are made at a spatial resolution of 0.25° latitude/longitude (approximately 25 km) and with a 6 hour timestep. It receives the two previous states of the atmosphere as input -- in other words, the atmospheric state at 0h and -6h is used to predict the state at +6h. A forecast is then made by iteratively calling GraphCast to step forward the atmospheric state. GraphCast is trained using a mean squared error loss. Training is autoregressive out to lead times of 3-days, to improve the model's ability to predict the atmospheric evolution over several timesteps. The loss function does not weight all variables equally: the single most important variable in the loss function is the surface 2m Temperature (2mT) field. This variable receives a weighting in the loss function equal to that of an entire 3D variable (on 37 pressure levels) such as temperature \citep{lam_learning_2023}. 

Graphcast is trained to emulate ERA5 (the ECMWF Reanalysis version 5) \citep{hersbach_era5_2020} using data from 1979--2017. ERA5 is based on the data assimilation system of the ECMWF Integrated Forecasting System (IFS) Cy41r2, which became operational on 8 March 2016. The reanalysis is produced using a 4D-Var data assimilation process, which blends observations taken over a period of 12 hours with short range (background) forecasts from the previous analysis update. In 1979, these observations numbered approximately 0.75 million per day increasing to around 24 million a day by 2019 \citep{hersbach_era5_2020}. ERA5 is produced at a horizontal resolution of 31 km, with 12-hour cycling from 9pm to 9am and vice versa, with hourly output saved. In addition to the main reanalysis, uncertainty information is generated by producing a lower resolution ten-member ensemble of 4D-Var reanalyses. This ensemble provides background error covariance estimates for the high-resolution 4D-Var computation.

In this study we choose to focus on 2mT because of the high weighting given to this field in the GraphCast loss function. This means that GraphCast assigns high importance to accurate forecasts of this field during training. Errors in this field therefore have particular significance when considering the possible limitations of GraphCast's forecasting ability. However, 2mT is not a prognostic variable in the IFS. It is diagnosed by interpolating between the temperature at the lowest model level and the temperature at the surface, using the same assumed profiles as in the surface flux calculations \citep{IFS_Physical_2024}. For this reason, ERA5 2mT is not directly analysed in the main 4D-Var computations. Instead it is estimated (alongside 2m relative humidity) through a two dimensional optimal interpolation (2D-OI) scheme which combines screen-level observations with the background forecast \citep{IFS_DataAssimilation_2024}. 

As described in \citep{IFS_DataAssimilation_2024}, during 2D-OI the background forecast at each analysis time (00, 06, 12, and 18 UTC) is first interpolated horizontally to the location of the observations. There is usually a mismatch between the altitude of the interpolated forecast and the true observation, because of the smoothed representation of orography in the forecast model: the observation is corrected to account for this discrepancy using a lapse rate of 4.5K/km. A quality control (QC) check is then carried out, and observations are rejected if they differ from the background forecast by more then 7.5°C. Background departure points, $\Delta X_i$, are calculated at each observation site, $i$, as the difference between each corrected observation and the interpolated background forecast. The final analysis increment, $\Delta X^a_p$, at each model grid point is computed as the weighted sum over the background departure points from the $N=50$ closest sites:
\begin{equation}
  \Delta X^a_p = \sum_{i=1}^N W_i \cdot \Delta X_i 
\end{equation}
where the weighting, $W_i$, is computed using the assumed background and observation error covariance matrices in a standard Kalman Filter update. In the weighting computation, the horizontal structure function has a decorrelation length scale set to 300 km, which results in a horizontal e-folding distance of 420 km, while the vertical structure function has a length scale of 800 m. If a given site has no valid observation at the time of interest, the departure point corresponding to the nearest in time (and most recent) valid observation within a plus/minus three-hour window is used instead. The standard deviations of background and observation errors are taken to be 1.5 K and 2 K respectively. 

We initialise GraphCast from ERA5 every six hours between 00:00UTC 1 January 2014 and 12:00UTC 31 Dec 2016. For each initial condition we produce forecasts for one time step, corresponding to a lead time of 6 hours. We compute the error in the GraphCast 2mT forecast as:
\begin{equation}
    \varepsilon = 2mT_{GraphCast} - 2mT_{ERA5}
\end{equation}
The recent period 2014-16 was chosen for our analysis to ensure the best observational data coverage and therefore constraint of ERA5. It also falls within the training period for GraphCast. This is to assess GraphCast's ability to emulate the behaviour of ERA5 as observed in the training data, as opposed to its ability to generalise to unseen data.


\section{Errors in 2m Temperature} \label{sec:error}

\begin{figure*}
    \centering
    \includegraphics[width=1\linewidth]{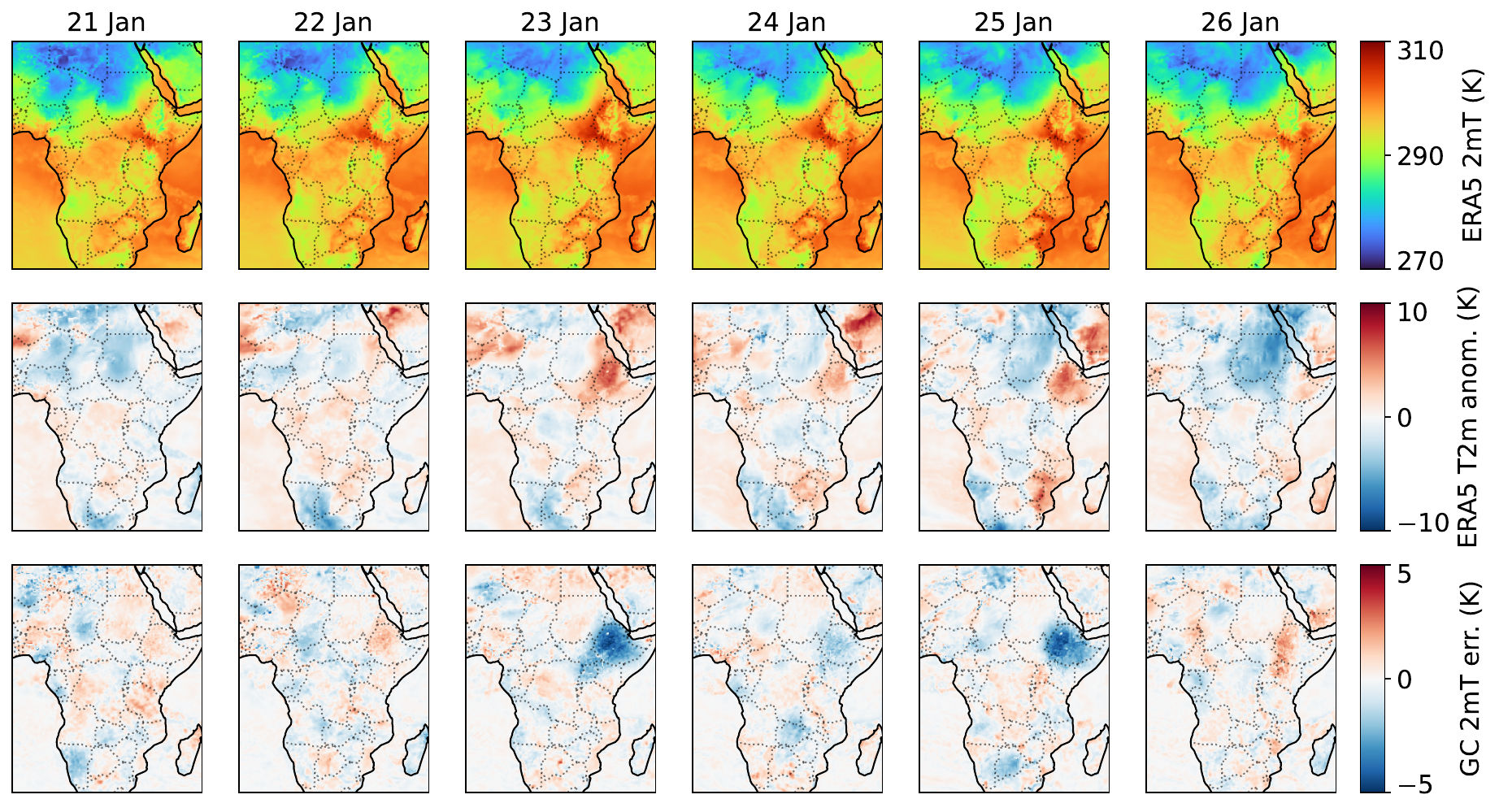}
    \caption{ERA5 2m Temperature (top row), ERA5 2m Temperature anomalies (centre row), and errors in 6h lead time GraphCast forecasts (bottom row), all shown at 06:00UTC for six consecutive days in 2016. ERA5 anomalies are computed by subtracting the 06:00UTC 2014-2016 January mean from the instantaneous fields. Error maps are defined as the GraphCast forecast validating at 06:00 UTC minus the corresponding ERA5 field.}
    \label{fig:stampplots}
\end{figure*}

To summarise the error database, we first performed an EOF analysis of the 6-hour GraphCast errors for forecasts validating at 06:00UTC. EOFs 3, 4 and 5 all showed that a strong component of 2mT error variability was localised over the Ethiopian Highlands in central and northern Ethiopia (Figure S1). This merited further investigation. We also performed the EOF analysis for forecasts validating at other times. Although they show some signal of an error component over Ethiopia, the signal is strongest at 06:00UTC.

Figure~\ref{fig:stampplots} (bottom row) shows a sequence of 06:00UTC error maps for six consecutive days in 2016. On 23rd and 25th January, a large cold error of up to -5K is observed in the GraphCast forecast. We call these `error events'. The error events are not observed on the other days in the period. The error is approximately centred on Addis Ababa, and is order 1,000 km in diameter. The top row of Figure~\ref{fig:stampplots} shows the ERA5 2mT fields at 06:00UTC for the same days. No unusual features are obvious on the days corresponding to the error events. The centre row shows ERA5 anomalies for the same days, computed by subtracting the 2014-2016 January 06:00UTC mean from each raw field. Here we observe that ERA5 is warmer than average over Ethiopia on the days with the error events, though the anomaly is not out of the ordinary compared to anomaly features in other regions. For example, there are strong warm anomalies over both Saudi Arabia and Mozambique on certain days within the selected period, but these other warm anomalies do not correspond to an error feature in GraphCast forecasts.

\begin{figure*}
    \centering
    \includegraphics[width=1\linewidth]{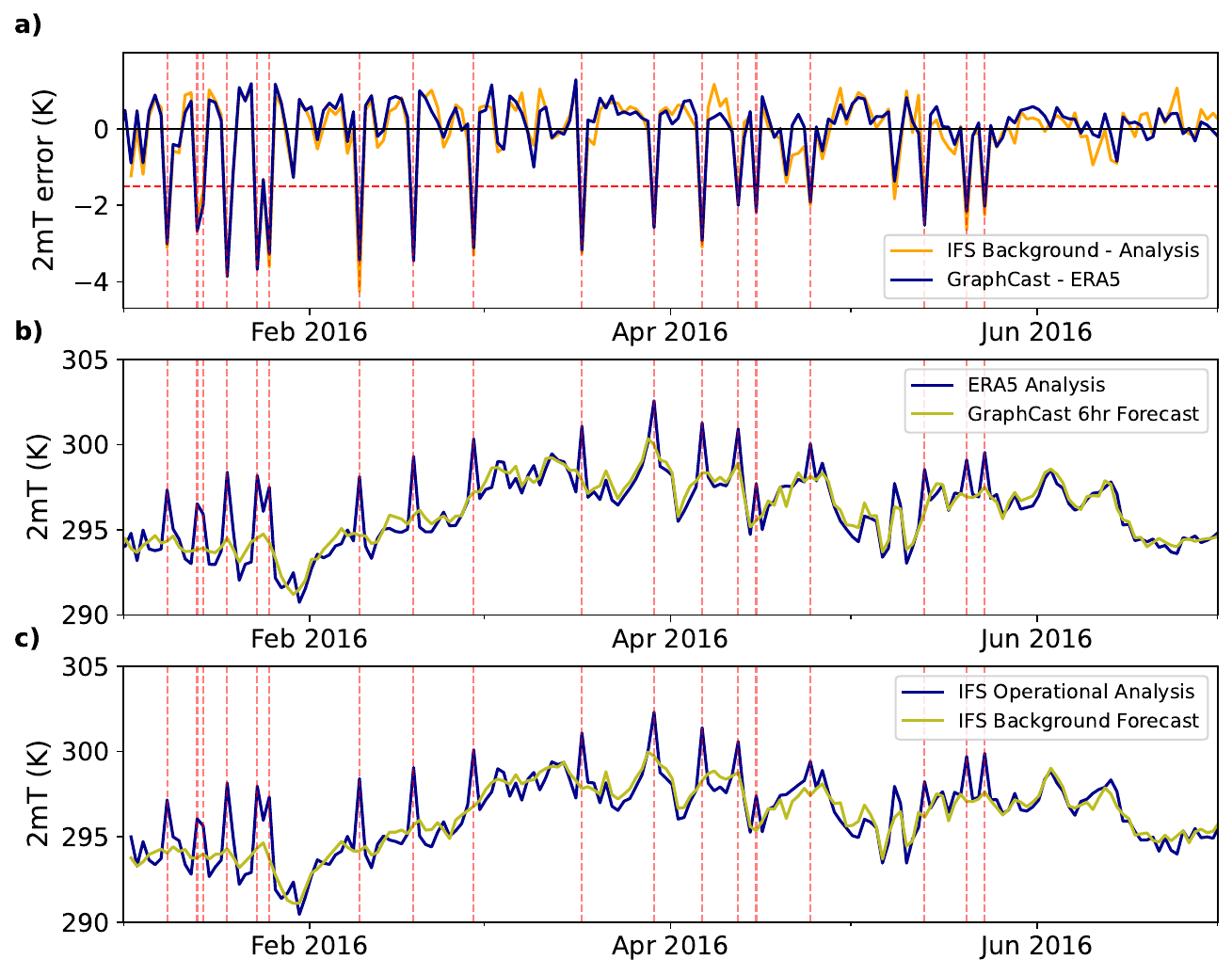}
    \caption{(a) Timeseries of the error in GraphCast compared to ERA5 (dark blue), and the difference between the background forecast and the analysis from within the ECMWF operational data assimilation system (yellow). (b) GraphCast 6h forecasts compared to ERA5. (c) ECMWF background forecasts and analysis from the operational data assimilation system. All panels show results for 2m Temperature, are averaged over 7.5-13.5°N, 35-41°E, and are valid at 06:00UTC on each date. The horizontal dashed line in panel (a) shows the threshold used to define error events in GraphCast compared to ERA5. The vertical dashed lines in each panel highlight the detected error cases.}
    \label{fig:timeseries}
\end{figure*}

We define a region of interest over Ethiopia spanning 7.5-13.5°N, 35-41°E, and compute the average error in this region. Figure~\ref{fig:timeseries}(a) shows a timeseries of this 06:00UTC error over the first half of 2016 (blue line). The error events are clearly visible. We plot the ERA5 analysis, and the GraphCast 6h forecasts in Figure~\ref{fig:timeseries}(b), demonstrating that it is ERA5 and not GraphCast which shows abrupt jumps for these error cases. Panel (c) plots the same timeseries for the ECMWF operational analysis and the corresponding background forecasts valid at 06:00UTC for each date, and the error in the operational analysis is included in panel (a). The ECMWF operational analysis shows similar behaviour to ERA5, while the corresponding operational background forecasts show similar behaviour to GraphCast. 

There are two possible causes of the observed error. Firstly, the error could point to a problem with GraphCast. For example, there could be a short-lived weather phenomenon that forecasting models struggle to predict, but that is captured by the reanalysis. We note that if this is the case, figure~\ref{fig:timeseries}(c) indicates the same problem in forecasts using the IFS. To identify what this phenomenon could be, we consider the distribution of variables in ERA5 on days with and without the error events. This includes skin reservoir content, total evaporation, cloud cover, various radiative fluxes, precipitation, and humidity and temperature at a range of vertical levels. We find no significant difference in distribution between error and non-error cases for any of these variables (not shown). We also assess whether related fields in ERA5 show similar fluctuations, including temperature on pressure levels between 900 hPa and 500 hPa, and skin temperature. These variables do not show the same unusual behaviour (Figure S2). We therefore conclude that the error is likely an issue with ERA5 that is also present in the operational analysis.

To compute statistics of the error, we define error events as points in time for which the GraphCast error averaged over the rectangular region 7.5-13.5°N, 35-41°E is below -1.5 K. This is shown as the horizontal dashed line in Figure~\ref{fig:timeseries}(a). A visual comparison of the maps of errors on those days indicates that this approach was effective at selecting error events, with very few false positives or missed events. Table~\ref{tab:stats} shows the number of error events at each GraphCast validation time for the 2014--16 period. To assess whether these errors are a recent phenomenon, or if they occur throughout the ERA5 record, we perform additional GraphCast forecasts to compute the statistics over 1980, 1990, 2000 and 2010 respectively. The results are shown in Table~\ref{tab:stats}. The error cases are strongly confined to 06:00UTC, and occur on around 7\% of dates. They are also present throughout the historical record. Maps of example error cases in the earlier part of the record are shown in Figures S3 and S4. For earlier years, the error cases over Ethiopia have a similar spatial structure. However, many additional error features are visible which could be worth investigating.

\begin{table}[h!]
\centering
\begin{tabular}{r r r r r}
\toprule
         & 00:00UTC & 06:00UTC & 12:00UTC & 18:00UTC \\
\midrule
1980     & 1.1 & 8.0 & 0.3 & 0.0 \\
1990     & 0.8 & 3.6 & 0.0 & 0.0 \\
2000     & 0.3 & 0.5 & 0.0 & 0.0 \\
2010     & 0.8 & 2.7 & 0.3 & 0.0 \\
2014--16 & 1.5 & 7.2 & 0.2  & 0.1 \\
\bottomrule
\end{tabular}
\caption{Percentage of ERA5 2m Temperature fields with error cases, as a function of time of day, for all dates within different reanalysis years. Here we define error events as points in time for which the GraphCast error averaged over the rectangular region 7.5-13.5°N, 35-41°E is below -1.5 K.}
\label{tab:stats}
\end{table}

The ERA5 ensemble of data assimilation (EDA) aims to represent uncertainty in the analysis due to the observing system, boundary conditions, and the forecast model. It primarily accounts for random observation and model uncertainties, as opposed to systematic errors, but it can still provide information about relative uncertainty in the reanalysis. To assess whether the spread of the EDA correctly indicates higher uncertainty in the reanalysis for the error cases, we compute spread-error scatter plots \citep{leutbecher2009} for two point locations.

\begin{figure}
    \centering
    \includegraphics[width=0.7\linewidth]{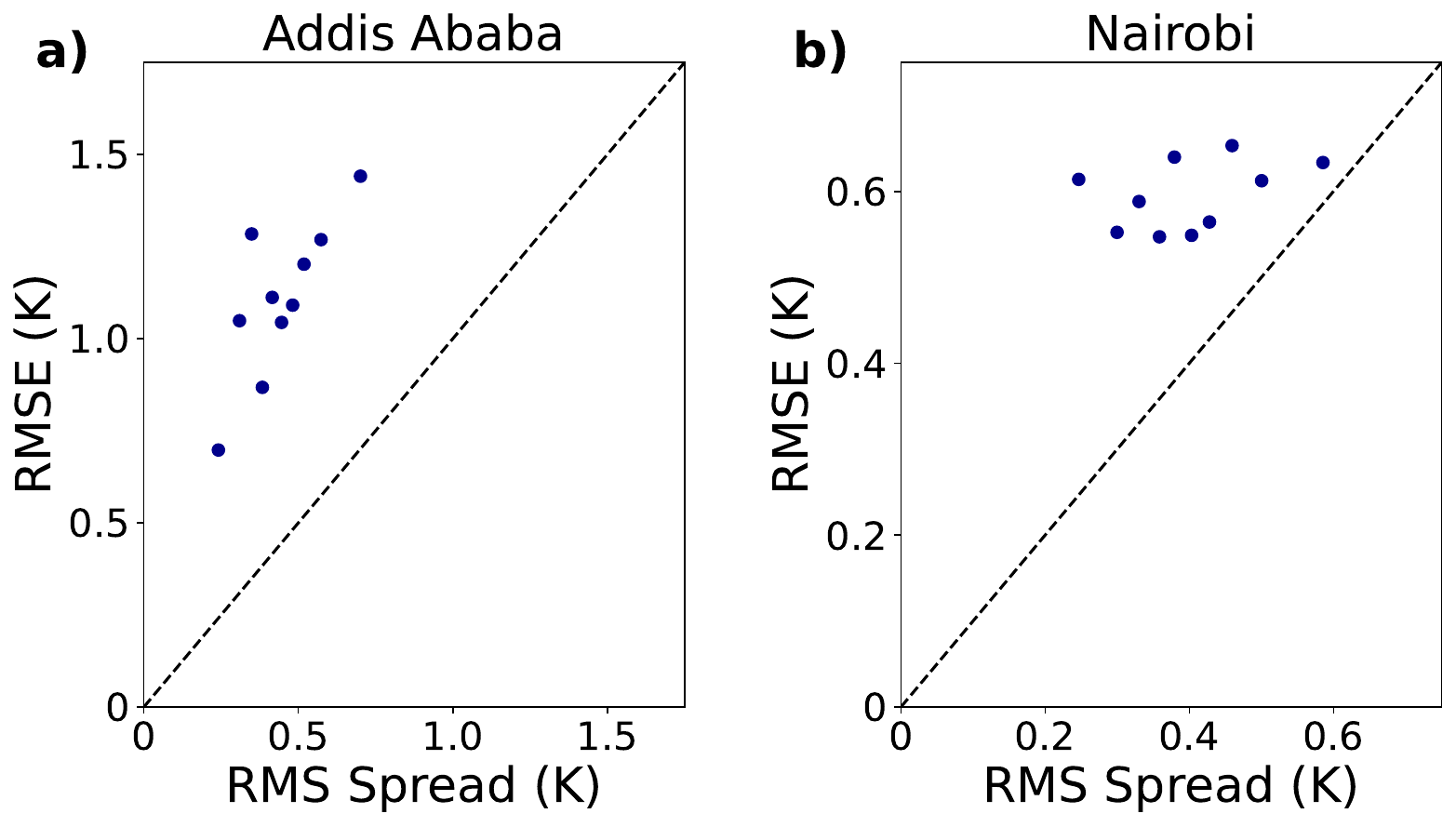}
    \caption{Root mean square (RMS) ensemble spread compared to RMS error for (a) Addis Ababa, Ethiopia (9°N, 38.75°E) and (b) Nairobi, Kenya  (1.25°S, 37°E). The ensemble spread is taken from the ERA5 ensemble of data assimilation whereas the error is computed between GraphCast and ERA5.}
    \label{fig:error_spread}
\end{figure}

All dates in 2014--16 are first sorted according to the EDA ensemble spread, and grouped into ten equally populated bins. Ideally, we would compare the EDA spread on each day to the reanalysis error, i.e. the difference between the reanalysis and the real world, but this is not available. We are primarily interested in whether the EDA spread captures uncertainty associated with the very large error cases. We therefore use GraphCast as a proxy for the real world, since for these large error cases the difference between GraphCast and ERA5 will be dominated by the reanalysis error (as opposed to the forecast error). Therefore, for each bin, the EDA root mean square (RMS) spread is computed, and compared to the RMS error in GraphCast forecasts for those corresponding dates. The results are shown in Figure~\ref{fig:error_spread}. The two locations were chosen as points close to international airports, to benefit from the automated screen-level observations available there. The spread-error plot for Addis Ababa, within our region, shows a clear relationship between EDA spread and GraphCast error. More error cases are typically found in high-spread bins than in low-spread bins, giving rise to differences in RMSE across those bins: the EDA spread provides useful information about the possibility of large error cases. However, the ensemble is underdispersive. The spread is approximately half the observed RMSE, and does not reach an amplitude that reflects the true error. In contrast, there is no clear relationship over Nairobi, which is outside of our region of interest. Errors between GraphCast and ERA5 at this location are likely dominated by random error growth as opposed to uncertainties in the reanalysis.

Finally we assess whether there is a systematically larger EDA spread associated with error cases compared to non-error cases. This can be answered by sorting the data by error instead of by spread, and binning as above. If we do this, the largest error bin has an RMS spread of 0.51 compared to its RMSE of 3.0. The spread of all other bins is approximately 0.45 and shows no relationship with RMS Error. So there is a systematic, albeit modest, increase in spread for error cases compared to non-error cases.


\section{The cause of the error} \label{sec:cause}

An analysis product represents the average weather conditions over the grid box, whereas a surface observation is a measurement of the local weather condition. This philosophy underpins the 2D-OI approach. A mismatch between the background forecast and the observation is to be expected, because of the difference between the grid-box average and local weather conditions. The QC threshold is therefore set high to ensure that valid observations are retained and used to constrain the analysis. To better represent the grid-box average and reduce computational costs, the 2D-OI scheme uses the nearest 50 observations to constrain the analysis at any given point. Regions with complex orography, such as Ethiopia, are particularly challenging. ERA5 has a horizontal resolution of 31km (corresponding to an effective resolution of $\sim$250km) which is not sufficient to resolve small-scale surface features, nor the resultant localised weather conditions. This means there is often a very substantial mismatch between a point observation and the corresponding first-guess forecast. Since microclimates can vary over short distances in such regions, point-wise observations are very different to the grid-box average, and averaging over many observations becomes critical. 

A further challenge in this region is that surface observations are relatively sparse. 
At each analysis time, 20--30 observations are typically used to constrain the analysis across the Ethiopian region of interest; an area of approximately 700 x 700 km.
For comparison, Figure S5 shows a typical global distribution of surface observations assimilated into the 06:00 UTC 2mT field, highlighting substantially more observations in other global regions. Because the nearest 50 observations are used to constrain the analysis at a given point, in regions where surface observations are sparse, a single observation can influence the analysis over a very large region, including at locations with very different local weather effects and associated model biases. 

Initially our assumption was that the reanalysis errors were due to sparsity of observations in space; however on further investigation, we identified that the ultimate source of the error is actually sparsity of observations in time. We found that a cluster of stations in Ethiopia have unusual reporting practices. These stations often miss reporting observations at 0600 UTC (9 am local time), but report instead at 0900 UTC (local noon). However, the 2D-OI scheme only has a background forecast valid at the synoptic time of 0600 UTC.  Since a report at 0900 UTC falls within the acceptable plus/minus three hour window, the background at 0600 UTC is compared to the raw measurement at 0900 UTC. The diurnal cycle shows very large increases in temperature over this window, so the observation at 0900 UTC is always substantially warmer than the background forecast at 0600 UTC, and is usually QC rejected. However, occasionally, it slips below the threshold, leading to a large jump in analysis temperature at 0600 UTC. Because a cluster of stations in the region has this reporting practice, and because of the overall sparsity of stations, this leads to a Gaussian-shaped uplift in the temperature over a large region centred on Ethiopia. In addition to a measurement at 0900 UTC, a measurement is also often available at 0300 UTC. However, the 2D-OI scheme always selects the nearest and \emph{most recent} (i.e. later) observation, such that the 0900 UTC measurement is preferentially selected, and the errors always have a positive sign.

\section{Implications for MLWP} \label{sec:implications}



Finally, we explore the implications of the ERA5 errors for GraphCast. Firstly, we assess whether the presence of the error in the training data has made forecasts in this region measurably worse compared to other regions. We hypothesise that GraphCast (and other MLWP models) might show systematically warmer 2m Temperature predictions over Ethiopia at 06:00UTC to hedge against the error. This would be unphysical, but would better match the training distribution.

\begin{figure}
    \centering
    \includegraphics[width=0.5\linewidth]{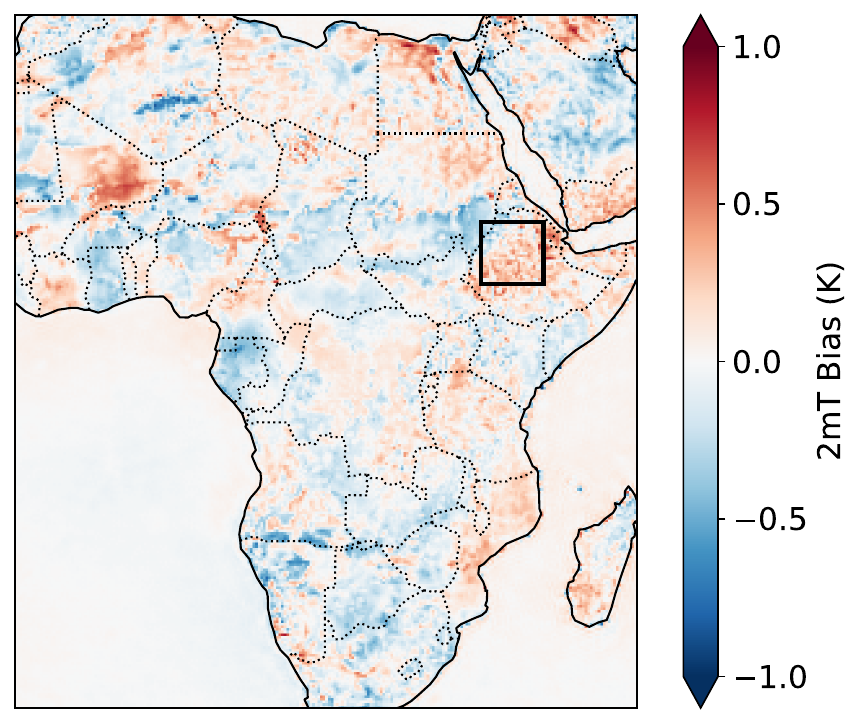}
    \caption{Bias in 6h GraphCast forecasts over Africa validating at 06:00UTC for non-error cases. The bias is computed over 2014-16, compared to ERA5.}
    \label{fig:bias_africa}
\end{figure}

\begin{figure}
    \centering
    \includegraphics[width=0.5\linewidth]{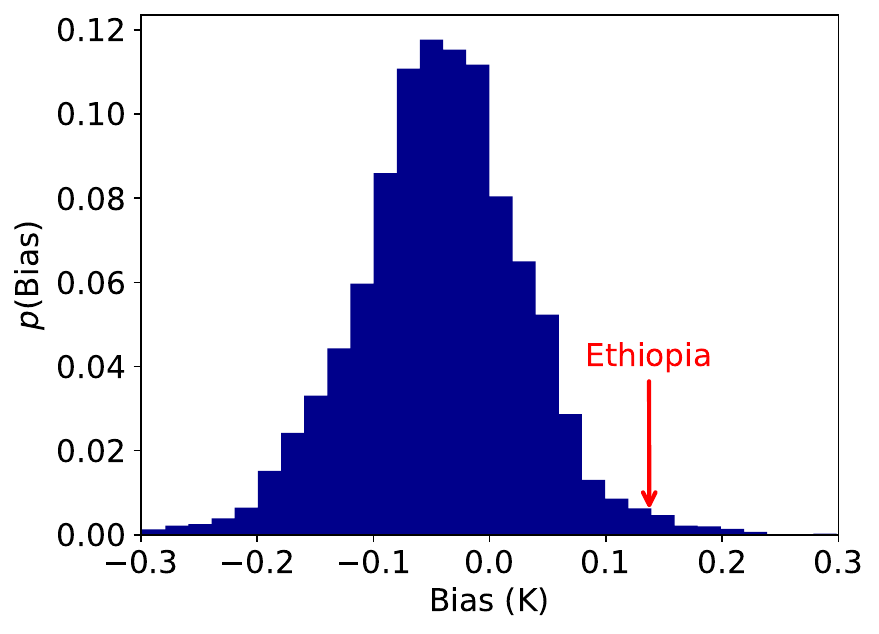}
    \caption{Distribution of regionally-averaged 2m Temperature biases in 6h GraphCast forecasts at 06:00UTC. The histogram contains only land regions sampled globally. The bias measured for the Ethiopian region is marked by the red arrow.}
    \label{fig:hist_biaspatch}
\end{figure}

We exclude all error cases from the database, leaving only forecasts where we are confident in the ERA5 validation field. Figure~\ref{fig:bias_africa} shows the 2mT GraphCast bias at 06:00UTC. Points within our region of interest have a slight warm bias. When averaged over our 6°x6° region (rectangle marked on Figure~\ref{fig:bias_africa}), the warm bias is +0.14K. However, other regions over Africa have similar magnitude biases. To assess the significance of the bias in the region of interest, we randomly sample 1 million 6°x6° boxes, where the lower left hand corners of these boxes are sampled from -90--84°N and 0--354°E. Sample boxes containing any ocean points are discarded, before the average 06:00UTC bias is computed for each of the remaining $\sim$186,000 samples. Figure~\ref{fig:hist_biaspatch} shows the distribution of the resultant regionally-averaged biases. GraphCast 2mT forecasts over land are, on average, slightly too cold at 06:00UTC. The Ethiopian region has an average bias of 0.14, which is 2.46 sigma from the mean. This corresponds to the 98.8th percentile, and is therefore a significant warm bias at the 5\% level, though we note that it is within the distribution of biases for regions of this size. Furthermore, an average bias of -0.14K falls at the 9.3rd percentile, such that biases of this \emph{magnitude} are not significant.

\begin{figure}
    \centering
    \includegraphics[width=0.7\linewidth]{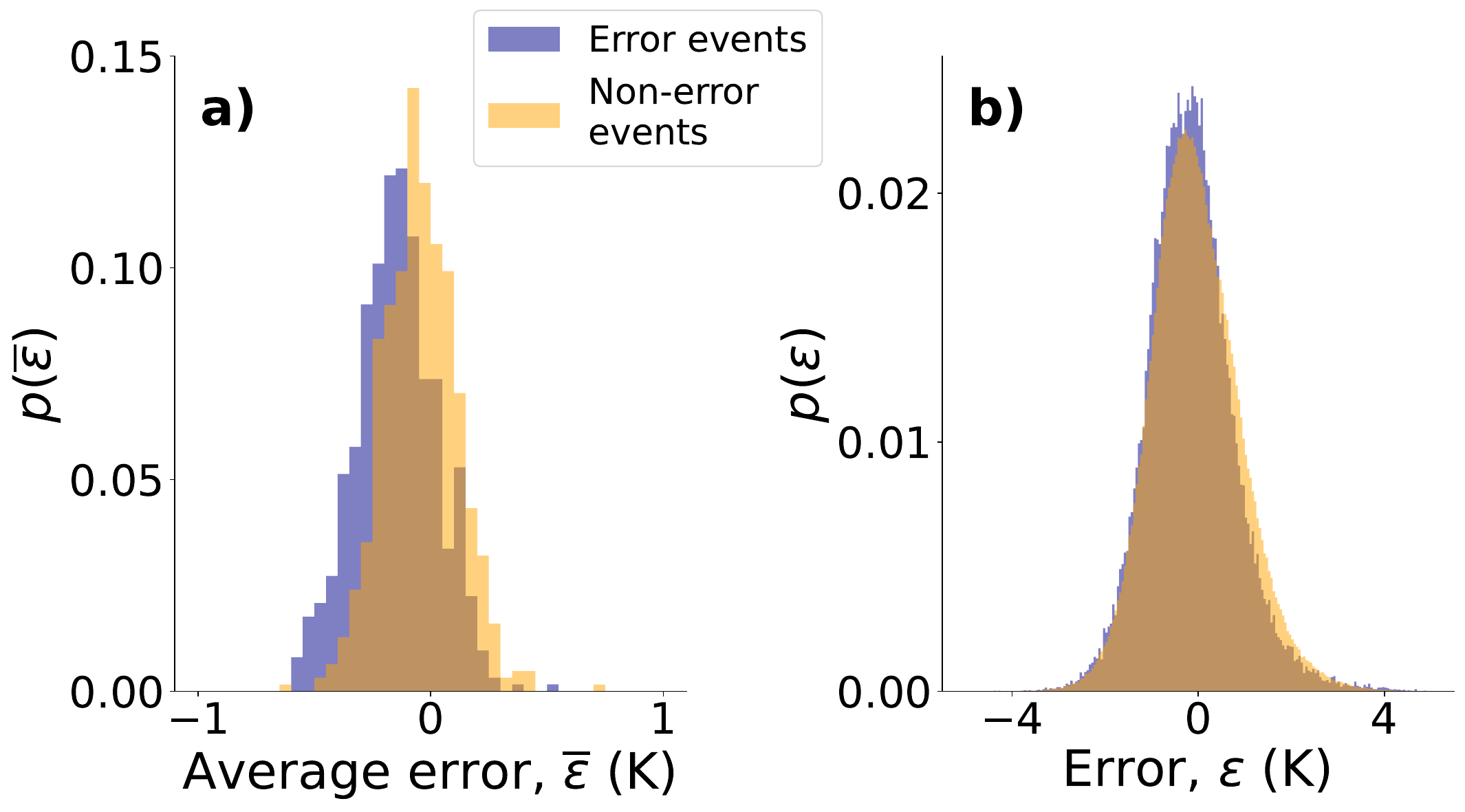}
    \caption{Distribution of 12:00UTC error in GraphCast forecasts initialised with (blue) and without (yellow) an error case in the initial conditions. a) Bias distribution, defined as error averaged in time over all start dates and b) instantaneous errors in both space and time. Both show data for all locations within the Ethiopian region of interest. }
    \label{fig:hist_bias_1200}
\end{figure}

Secondly, we assess whether the presence of an error-case in the initial conditions systematically impacts forecast skill at subsequent times. We sort the data into two groups according to the presence of an error case at 06:00UTC. We then compute the forecast bias at 12:00UTC for each group of forecasts, averaged over all initial conditions within the group, separately for each location. Figure~\ref{fig:hist_bias_1200}(a) shows the distribution of biases for each point within the region of interest over Ethiopia, for the initial conditions with and without the error cases. If GraphCast is able to ignore the warm error in its starting conditions, the bias distributions would be indistinguishable on error and non-error days. Instead, the forecasts initialised with the error cases show a systematic cold bias (average: -0.15K) at 12:00UTC compared to forecasts initialised without the error cases (average: -0.045K). GraphCast forecasts are therefore systematically colder for initial conditions with a warm error. However, we note that while there is a systematic shift, it is small compared to the size of the error in the initial conditions, which is typically $\sim$-3K (see Figure~\ref{fig:timeseries}). Figure~\ref{fig:hist_bias_1200}(b) shows the distribution of instantaneous errors, in both space and time, for these two groups of forecasts. This further demonstrates that the impact is small.


\section{Discussion and Conclusions}\label{sec:concs}

Here we summarise and expand the principal conclusions of this study, and then offer practical recommendations going forward.

\vskip10pt
\noindent \textbf{The error is significant because 2m Temperature is widely used.}

\noindent The 2mT field is one of the most widely downloaded ERA5 fields. It is commonly used for impact studies, climate change trend analysis, and verification of weather and climate models. The error we identify over the Ethiopian highlands is pervasive throughout the dataset and occurs at 0600 UTC on an appreciable fraction of dates (approximately 7\% in our sample). Consideration of samples from 1980 and 1990 indicate this may be a wider-spread issue in the earlier part of ERA5, with more geographical regions affected. It is important to report this error to users of ERA5 so they can account for it in any use of ERA5 for impacts or model evaluation.

\vskip10pt

\noindent \textbf{Uncertainty estimates from ERA5 should be better utilised.}

\noindent Reanalysis products such as ERA5 blend observations with background forecasts from a numerical model. They inherit key benefits, but also limitations, from both these data sources. ERA5 includes an ensemble of data assimilation (EDA), which produces an uncertainty estimate used within the DA procedure, and which accounts for both model and observation error \citep{hersbach_era5_2020}. This uncertainty estimate is also available to users of the reanalysis product. Despite its availability, this uncertainty estimate is not routinely used. For example, it was not included with WeatherBench \citep{rasp2020weatherbench,Rasp2024}, and is not widely used by the MLWP community. Where possible, users and machine learning (ML) practitioners should make explicit use of the EDA (or other uncertainty estimates) to down-weight questionable analyses during training and verification, for example, through a including analysis uncertainty in the loss function. Furthermore, a better understanding of the DA process could guide how MLWP training objectives are constructed. For example, weighting surface 2mT equally with the 3D prognostic variables benefitted GraphCast’s overall skill, but surface fields may contain artefacts due to the 2D-OI algorithm and sparse surface observations, whereas 3D fields are relatively well constrained by 4DVar. 

\vskip10pt

\noindent \textbf{MLWP in the tropics requires caution because of observation sparsity.}
\noindent MLWP methods are often promoted as promising for tropical forecasting where NWP models can have lower skill. However, our study indicates that caution must be taken. Tropical regions are often more sparsely observed than the extra tropics, 
and remote or complex terrain amplifies the impact of atypical reporting practices or DA choices. However, the sparsity of surface observations means that end-to-end MLWP, which takes observations as input and predicts the forecast at station locations, would leave many users without a forecast. We must therefore assess the observing system and DA behaviour in regions of interest before relying on MLWP outputs for operational or impact use.

\vskip10pt

\noindent \textbf{MLWP can learn climatology shortcuts as well as physical relationships.}
\noindent Machine-learnt emulators are, by construction, statistical approximators of their training distribution. When training data contains intermittent errors, ideally an emulator could learn to ignore those samples. However, in practice it is likely that the emulator will either learn to reproduce those errors or hedge against them by shifting their predictions toward the climatological mean. Both behaviours are undesirable. 
Our analysis shows GraphCast exhibits a small but significant regional warm bias at 06:00 UTC consistent with the training distribution, and initial conditions containing the ERA5 error produce slightly different subsequent forecasts. This is evidence that ML models do respond to, and partially internalise, such errors. However, the biases are small. In the context of MLWP, there is substantial discussion as to whether such models have `learnt the underlying physics'. In our context we phrase this as: can an MLWP model identify that, on certain days, the 2mT field is inconsistent with the remaining 3D temperature fields? On the whole our results show that GraphCast has largely ignored the error, despite its prevalence throughout the training data. We note that this is despite the fact that the error is not purely random --- it is systematically more likely to occur on days with a smaller diurnal cycle. This hints that GraphCast has been able to identify and ignore this inconsistency to some extent.

\vskip10pt

\noindent \textbf{MLWP models can be valuable for assessing reanalysis quality.}

\noindent An unexpected positive outcome of this work is methodological: we found that comparing MLWP short-lead time forecasts to their training analysis was an effective way to detect intermittent errors in the reanalysis. Because ML models learn the dominant spatio-temporal structure of their training data, sudden, localised discrepancies between a forecast and the analysis can highlight errors that would otherwise be difficult to detect by eye. We therefore recommend using MLWP verification as a complementary diagnostic in reanalysis validation workflows, for those for whom the background forecast is not available. While NWP models could also be used to detect such errors, the portability and speed of MLWP models makes them particularly suitable for this task.

\vskip10pt

\noindent \textbf{Practical recommendations for reanalysis practice}

\noindent Finally, we consider recommendations for improving the analysis and reanalysis in this region, and for other locations in space and time which suffer from sparsity of observations and unusual reporting practices. In the 2D-OI system currently used to produce the 2mT analyses, we suggest that it is important to make better use of observations in time. The natural solution is to produce and use a background forecast that has a high time resolution, such that observations could be compared to the background that is valid at the time that the observation was made, as in 4DVar. In addition, instead of using only the nearest and most recent observation, it would be better to 1) use observations symmetrically in time, e.g. using an observation at both 0300 UTC and 0900 UTC should 0600 UTC be unavailable; and 2) to moreover consider using all observations available within the 6 hr window, suitably weighted according to their distance in time from the target time. These two changes should enable the system to perform better in regions with a rapidly changing diurnal cycle, as it would no longer be systematically more likely to look ahead than behind to find a suitable observation. In addition, making use of more observations in time will reduce random errors associated with individual observation locations, and also reduce jumpy behaviour.
A more fundamental solution would be to let 4DVar produce the 2mT analyses. This would allow a better and more extensive use of all available 2mT conventional observations in the assimilation window, in a manner consistent with the large number of other observations sensitive to near surface conditions. Work in this direction is ongoing.

\subsection*{Author contributions}
HC conceived the study and wrote the draft manuscript. JB performed analysis and made figures. BA created the error database. MB, MD and PdR discovered the source of the error. All authors discussed and interpreted the results and improved the final paper. 

\subsection*{Acknowledgments}
HMC and BA were supported through the EERIE project (Grant Agreement No 101081383) funded by the European Union. Views and opinions expressed are however those of the author(s) only and do not necessarily reflect those of the European Union or the European Climate Infrastructure and Environment Executive Agency (CINEA). Neither the European Union nor the granting authority can be held responsible for them. University of Oxford's contribution to EERIE is funded by UK Research and Innovation (UKRI) under the UK government’s Horizon Europe funding guarantee (grant number 10049639). HMC was also funded  through a Leverhulme Trust Research Leadership Award.

\subsection*{Data Availability} 
The ERA5 datasets used in the current study are available in the Copernicus Climate Change Service, Climate Data Store, (2023), for registered users. The following datasets were used: ERA5 hourly data on pressure levels from 1940 to present (DOI: 10.24381/cds.bd0915c6), ERA5 hourly data on single levels from 1940 to present (DOI: 10.24381/cds.adbb2d47). The IFS operational background and analysis datasets are taken from the long window 4DVar dataset (stream=lwda) in the ECMWF MARS archive \citep{mars_2018}.

\subsection*{Financial disclosure}

None reported.

\subsection*{Conflict of interest}

The authors declare no potential conflict of interests.

\bibliographystyle{wileyNJD-Harvard}
\bibliography{Christensen,journals}

\clearpage

\section*{Supporting information}

Additional supporting information may be found in the
online version of the article at the publisher’s website: Supplementary Figure 1: First 10 Empirical Orthogonal Functions (EOFs) of GraphCast 2mT errors.
Supplementary Figure 2: Timeseries of additional ERA5 fields averaged over the Ethiopian region, for the same time period as Figure 2.
Supplementary Figure 3: Errors in the 6hr lead time GraphCast 2mT forecasts validating at 06:00UTC, for 24 error cases in 1980.
Supplementary Figure 4: Errors in the 6hr lead time GraphCast 2mT forecasts validating at 06:00UTC, for 12 error cases in 1990.
Supplementary Figure 5: Spatial distribution of surface observations assimilated into the ECMWF 06:00 UTC operational analysis on 20 November 2025.

\clearpage

\begin{figure*}
    \centering
    \includegraphics[width=\linewidth]{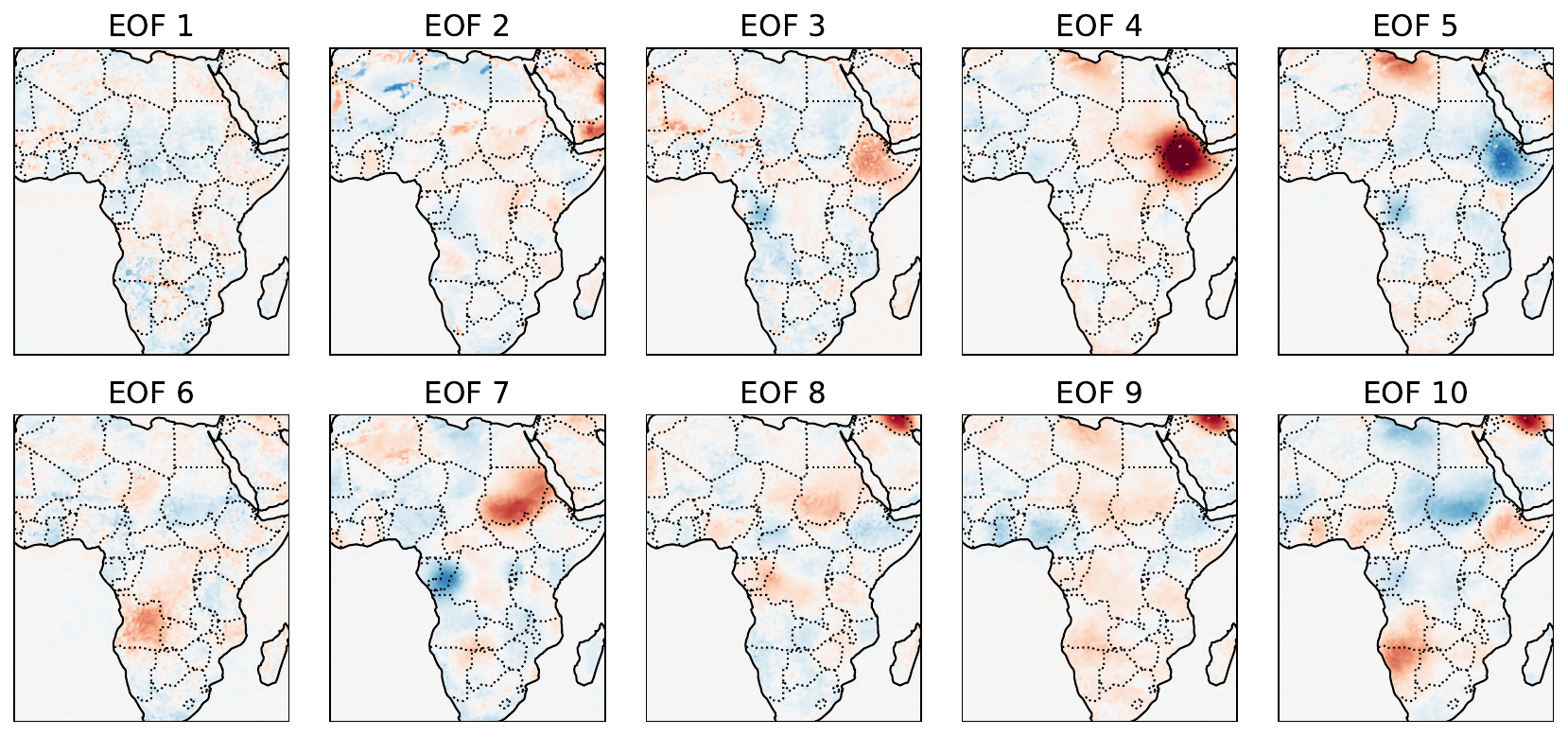}
    \caption*{Figure S1: First 10 Empirical Orthogonal Functions (EOFs) of GraphCast 2mT errors. The colour scale is arbitrary, and symmetric about zero. The EOFs are calculated on a dataset consisting of GraphCast errors at 06:00UTC for dates between 1 January 2014 and 31 December 2016. EOFs are computed using global data fields, though the leading EOFs are only shown over Africa.}
    \label{fig:SM1}
\end{figure*}

\begin{figure*}
    \centering
    \includegraphics[width=\linewidth]{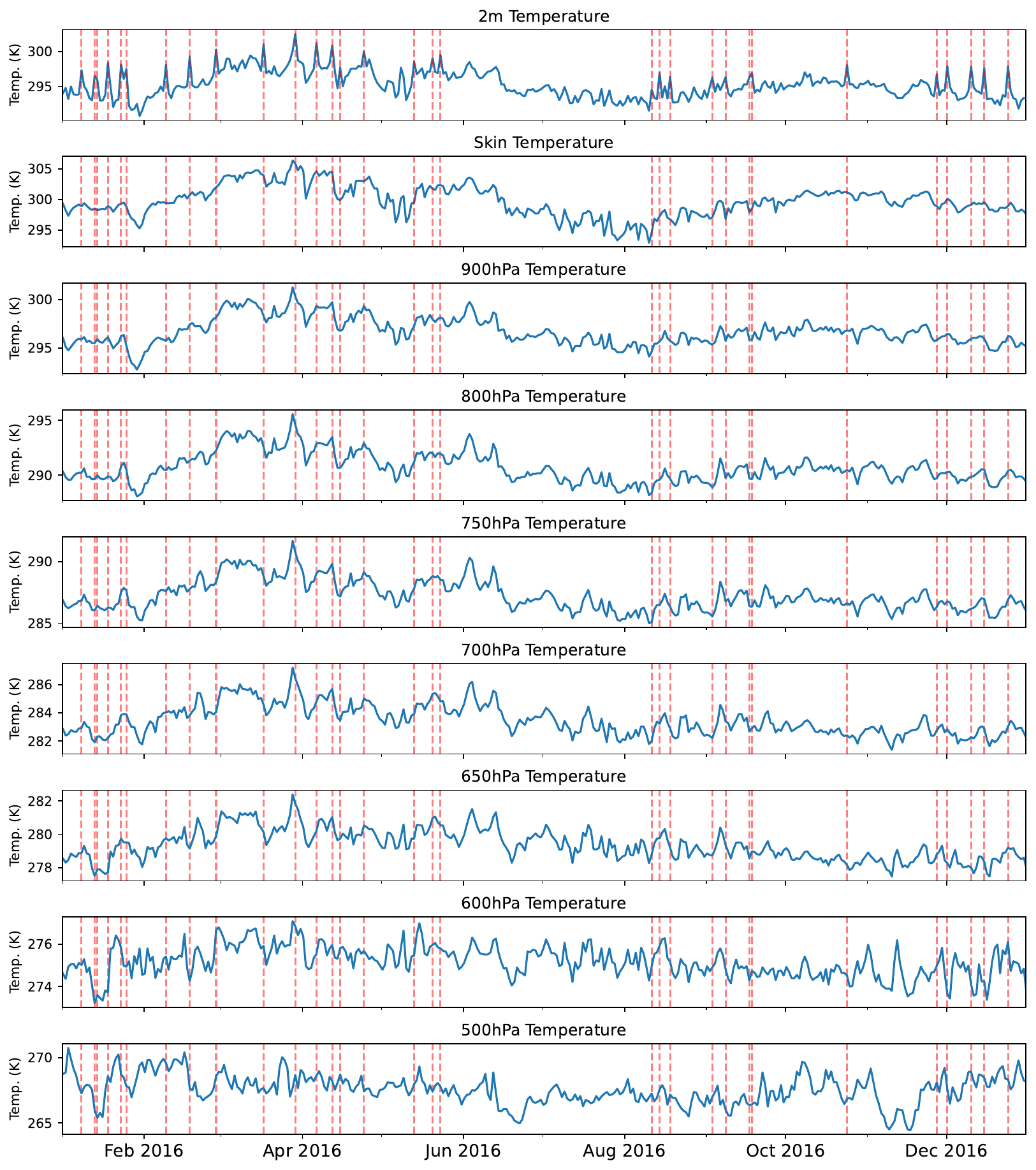}
    \caption*{Figure S2: Timeseries for a selection of ERA5 temperature fields, averaged over the region 7.5-13.5°N, 35-41°E at 06:00UTC. The vertical dashed lines highlight the 2mT error cases detected by the method described in the main text. The large fluctuations are only observed for 2m Temperature, and not for any other Temperature field.}
    \label{fig:SM2}
\end{figure*}

\begin{figure*}
    \centering
    \includegraphics[width=1\linewidth]{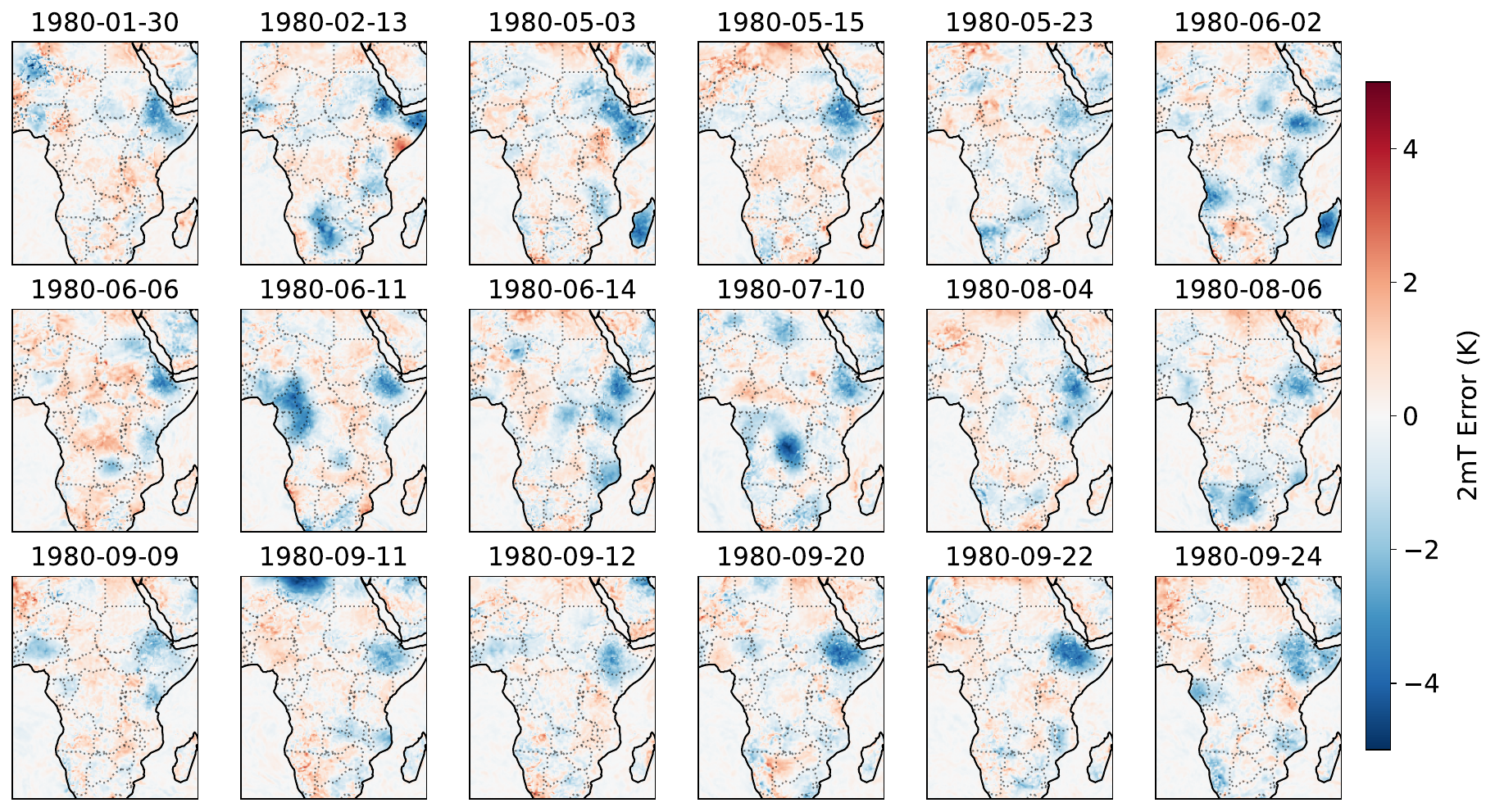}
    \caption*{Figure S3: Errors in the 6hr lead time GraphCast 2mT forecasts validating at 06:00UTC. The first 24 error cases in 1980 identified by the method described in the main text are shown (out of a total of 29).}
    \label{fig:SM4}
\end{figure*}

\begin{figure*}
    \centering
    \includegraphics[width=1\linewidth]{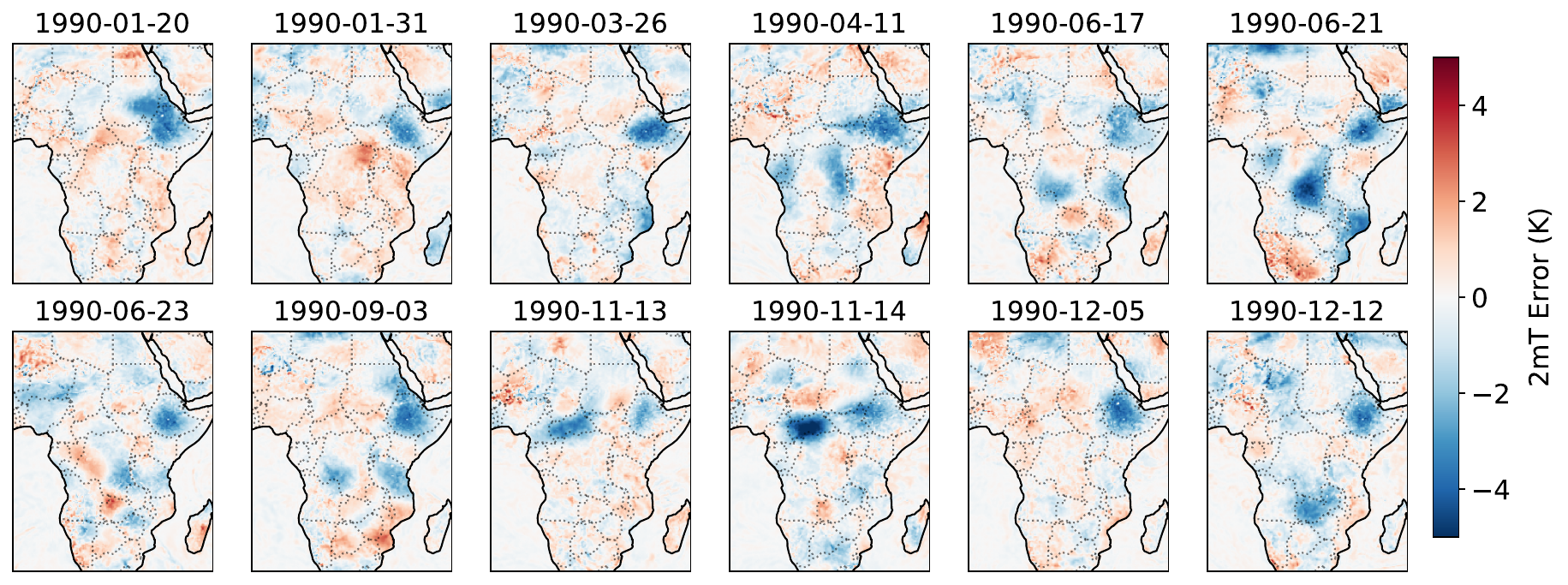}
    \caption*{Figure S4: Errors in the 6hr lead time GraphCast 2mT forecasts validating at  06:00UTC. The first 12 error cases in 1990 identified by the method described in the main text are shown (out of a total of 13).}
    \label{fig:SM5}
\end{figure*}

\begin{figure*}
    \centering
    \includegraphics[width=1\linewidth]{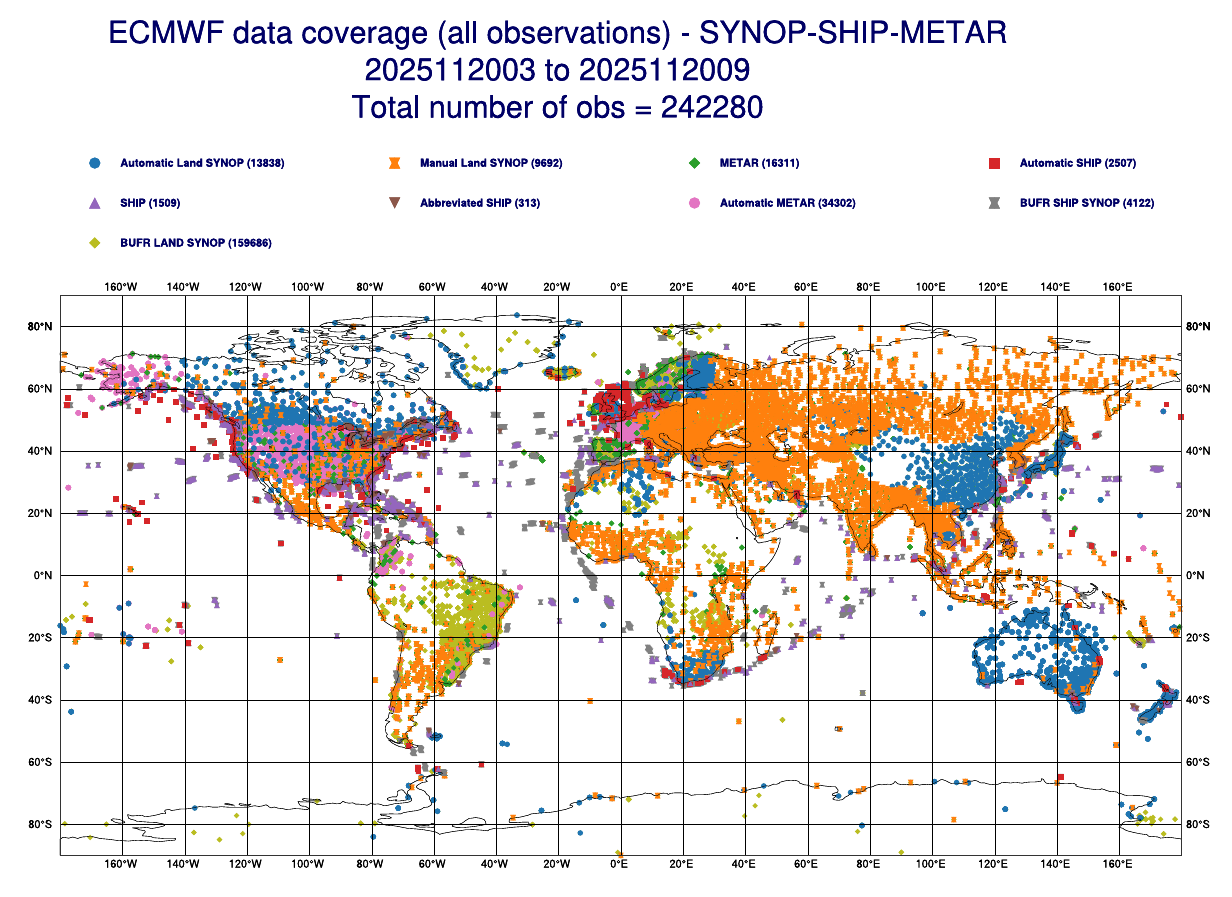}
    \caption*{Figure S5: Spatial distribution of surface observations assimilated into the ECMWF 06:00 UTC operational analysis on 20 November 2025.}
    \label{fig:obs_globe}
\end{figure*}

\end{document}